# Automatic linear measurements of the fetal brain on MRI with deep neural networks

Netanell Avisdris[1,2], Bossmat Yehuda[2,3], Ori Ben-Zvi[2,3], Daphna Link-Sourani PhD[2],
Dr. Liat Ben-Sira MD[3,4,5], Dr. Elka Miller MD[6], Dr. Elena Zharkov MD[7],
Prof. Dafna Ben Bashat PhD[2,3,5], Prof. Leo Joskowicz PhD[1]

1 School of Computer Science and Engineering, The Hebrew University of Jerusalem, Israel
2 Sagol Brain Institute, Tel Aviv Sourasky Medical Center, Israel
3 Sagol School of Neuroscience, Tel Aviv University, Israel
4 Division of Pediatric Radiology, Tel Aviv Sourasky Medical Center, Israel
5 Sackler Faculty of Medicine, Tel Aviv University, Israel
6 Medical Imaging, Children's Hospital of Eastern Ontario, University of Ottawa, Canada
7 Radiology, Shaare Zedek Medical Center, Israel

**Corresponding author**: Netanell Avisdris; **email**: *netana03@cs.huji.ac.il*

**Acknowledgments**: This research was supported by Kamin grant 63418 from the Israel Innovation Authority; Joseph Bar Nathan Trustee of Mychor Trust Fund. We are grateful to Vicki Myers for editorial assistance, and MRI technicians for scanning the fetuses.

Abstract

**Purpose:** Timely, accurate and reliable assessment of fetal brain development is essential to reduce short and long-term risks to fetus and mother. Fetal MRI is increasingly used for fetal brain assessment. Three key biometric linear measurements important for fetal brain evaluation are Cerebral Biparietal Diameter (CBD), Bone Biparietal Diameter (BBD), and Trans-Cerebellum Diameter (TCD), obtained manually by expert radiologists on reference slices, which is time consuming and prone to human error. The aim of this study was to develop a fully automatic method computing the CBD, BBD and TCD measurements from fetal brain MRI.

**Methods:** The input is fetal brain MRI volumes which may include the fetal body and the mother's abdomen. The outputs are the measurement values and reference slices on which the measurements were computed. The method, which follows the manual measurements principle, consists of five stages: 1) computation of a Region Of Interest that includes the fetal brain with an anisotropic 3D U-Net classifier; 2) reference slice selection with a Convolutional Neural Network; 3) slice-wise fetal brain structures segmentation with a multiclass U-Net classifier; 4) computation of the fetal brain midsagittal line and fetal brain orientation, and; 5) computation of the measurements.

**Results:** Experimental results on 214 volumes for CBD, BBD and TCD measurements yielded a mean $L_1$ difference of 1.55mm, 1.45mm and 1.23mm respectively, and a Bland-Altman 95% confidence interval ($CI_{95}$) of 3.92mm, 3.98mm and 2.25mm respectively. These results are similar to the manual inter-observer variability, and are consistent across gestational ages and brain conditions.

**Conclusions:** The proposed automatic method for computing biometric linear measurements of the fetal brain from MR imaging achieves human level performance. It has the potential of being a useful method for the assessment of fetal brain biometry in normal and pathological cases, and of improving routine clinical practice.

1. Introduction

Human fetal brain development is a complex process that involves significant changes in volume, structure, and maturation in a unique spatio-temporal pattern. Abnormal fetal brain development can have significant short and long-term consequences on the newborn. Consequently, accurate quantitative assessment of fetal brain growth is essential for early diagnosis of developmental disorders.

Ultrasound (US) is currently the primary imaging modality to monitor fetal development. Magnetic Resonance Imaging (MRI) is increasingly used for fetal brain assessment in cases of inconclusive US findings, to confirm or reject suspected abnormalities, and to detect other developmental abnormalities. MRI-based routine clinical assessment of fetal brain development is mainly subjective, with a few biometric linear measurements. Similar to US-based evaluation, these measurements are compared to MRI reference of growth centiles of normal developing fetuses [1,2].

Three key biometric linear measurements used in routine clinical assessment on fetal brain MRI are Cerebral Biparietal Diameter (CBD), Bone Biparietal Diameter (BBD), and Trans Cerebellum Diameter (TCD). These measures are performed manually on individual MRI reference slices by clinicians following established guidelines [3,4], which differ from the guidelines for US-based measurements, specify how to establish the scanning imaging plane, how to select the reference slice in this volume for each measurement, and how to identify the two anatomical landmarks for the linear measurement. The CBD and BBD measurements are performed on the same slice, and are drawn perpendicular to the mid-sagittal line (MSL). The TCD is measured on a different reference slice by selecting the two antipodal landmark points on the fetal brain cerebellum contour, giving the diameter of the cerebellum.

Manual measurements require clinician training, are time consuming, and suffer from intra- and inter observer variabilities [5]. Since fetal brain measurements are small, i.e., 30-100mm, especially at early gestational age, small measurement errors may cause a significant shift in the corresponding fetal growth centile, leading to misdiagnosis and misguided pregnancy management [6].

Developing automatic methods for computing biometric fetal brain measurements presents numerous technical challenges. First, the method should follow the guidelines and steps explicitly and implicitly performed by the clinician, i.e., localization of the fetal brain in the MRI volume, selection of the reference slice, identification of the fetal brain, skull and cerebellum contours and mid-sagittal line, and selection of anatomical landmarks for each linear measurement. Each of these stages presents unique and significant computational challenges. Additional challenges include the variability of the MRI scanning planes, resolutions, contrasts and protocols, pathological fetal brain conditions, and fetal motion artifacts, all of which may affect image quality, and yield inaccurate measurements and significant observer variability.

In this paper, we present the first fully automatic method for computing three key biometric linear fetal brain measurements in MRI, i.e., CBD, BBD, and TCD.

2. Related work

To the best of our knowledge, there are no published reports of automatic biometric linear measurement methods of the fetal brain MRI. However, a variety of methods have been reported that are relevant to the five stages of our method. We review them next.

*Fetal brain ROI detection and segmentation:* Torrents-Barrena et al. [7] presented a comprehensive review of methods for segmentation and classification of fetal structures on US and MRI. Dudovitch et al. [8] presented a method for fetal brain ROI detection and segmentation based on two 3D U-Nets: one for ROI localization and one for ROI voxel classification. We used the fetal brain ROI localization 3D U-Net as the first stage of our method.

*Reference slice selection:* Baumgartner et al. [9] described a real-time CNN-based fetal US slice selection method, focusing on the temporal aspect of the reference slice selection, which differ from our problem as there is more than one reference slice solution. Pallenberg et al. [10] described a template-based spatial slice selection for CT.

However, this method cannot handle the significant variability in fetal brain morphology with gestational age. Various methods have been described for the selection of planes and computation of fetal linear measurements in 3D US scans. Li et al. [11] describe a method to compute the trans ventricular (TV) and the trans cerebellum (TC) planes, which are similar to the CBD/BBD and the TCD slice planes in MRI, respectively. Ryou et al [12] describe a method for selecting planes and computing the crown-rump-length (CRL), head circumference (HC) and abdomen circumference (AC). However, these problems differ from ours in that 3D US provides dense, isotropic spatial information while the fetal MRI is sparser and has different noise and sampling characteristics. In addition, the fetal MRI scan may have spatial motion artifacts that hamper the reconstruction of a spatial volume from planes.

*Fetal brain component segmentation:* Despotović et al. [13] presented a survey of model-based methods for segmentation of adult brain components in MRI. Most methods, e.g. FreeSurfer [14], use a brain atlas and require registration. They are not directly applicable to fetal brain scans since the fetal brain size, shape, and structure changes rapidly during gestation. Others have developed atlases for the various gestational ages for segmenting brain structures [15]. These methods require accurate 3D non-rigid registration, which is time-consuming and may be inaccurate [16]. More recently, deep learning methods have been developed for the segmentation of fetal brain structures [17] in fetal MRI scans. However, this method is not applicable to the problem at hand since it does not differentiate between the left and right hemispheres.

*Mid-sagittal line (MSL) computation:* to the best of our knowledge, there are no papers in the literature that describe methods for fetal brain mid-sagittal line computation. However, two types of methods for the computation of the adult brain MSL in MRI scans have been developed [18]: shape-based methods [19] and content-based methods in which the MSL is computed from the line that maximizes the brain's bilateral symmetry [20]. These methods are designed for T1-weighted adult brain MRI scans and rely on skull stripping prior to segmentation. This task is more challenging on T2-weighted fetal brain scans, as the fetal skull contrast is different and its boundaries are fuzzy.

To summarize, while biometric linear measurements of the fetal brain are an essential part of fetal development assessment, they are currently performed manually. While automatic methods for the computation of US-based biometric linear measurements are available, e.g., biparietal diameter [21,22], fetal head circumference [23] and femur length [21], no such methods are available for fetal MRI.

## 3. Method

We present a fully automatic method to compute three key fetal biometric measurements, CBD, BBD and TCD, from fetal brain MRI. The input is a fetal MRI volume. The outputs are the measurements and reference slices in which the measurements were computed. The method follows the clinical guidelines for manual fetal MRI measurements [4]. The pipeline consists of five stages (Fig. 1): 1) computation of a Region of Interest (ROI) of the fetal brain with an anisotropic 3D U-Net classifier; 2) reference slice selection with a convolutional neural network (CNN); 3) slice-wise fetal brain structure segmentation with a multiclass U-Net classifier; 4) computation of the fetal brain MSL and fetal brain orientation, and; 5) computation of CBD, BBD and TCD measurements. The method performs self-assessment of reliability and alerts clinicians when the measurements may be unreliable.

Our method relies on supervised deep learning techniques for the first three stages, which requires offline training and online inference (Fig. 2). In the offline training phase, the networks of the first three stages are trained individually on annotated training and validation datasets. In the online inference phase, these networks are used for inference. For the reference slice selection (stage 2), the CNN is trained twice, one to select the reference slice for CBD/BBD measurements and one for TCD measurement.

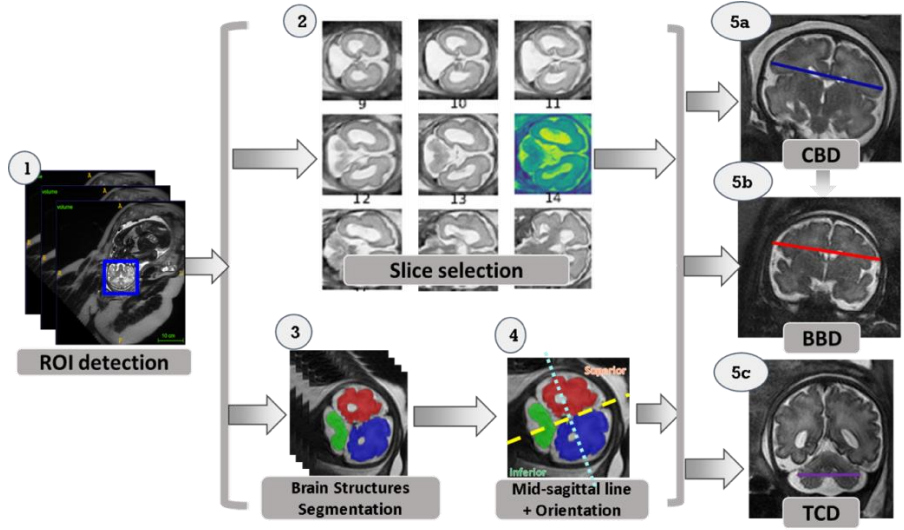

**Fig. 1**: Fetal MRI biometric measurements method: 1) fetal brain ROI detection (blue box); 2) reference slice selection (coronal slices, selected slice in yellow); 3) fetal brain structures segmentation: cerebellum (green), left (blue) and right (red) hemispheres; 4) MSL (yellow line) and brain orientation (separated by light blue line) computation; 5) CBD, BBD, and TCD measurements.

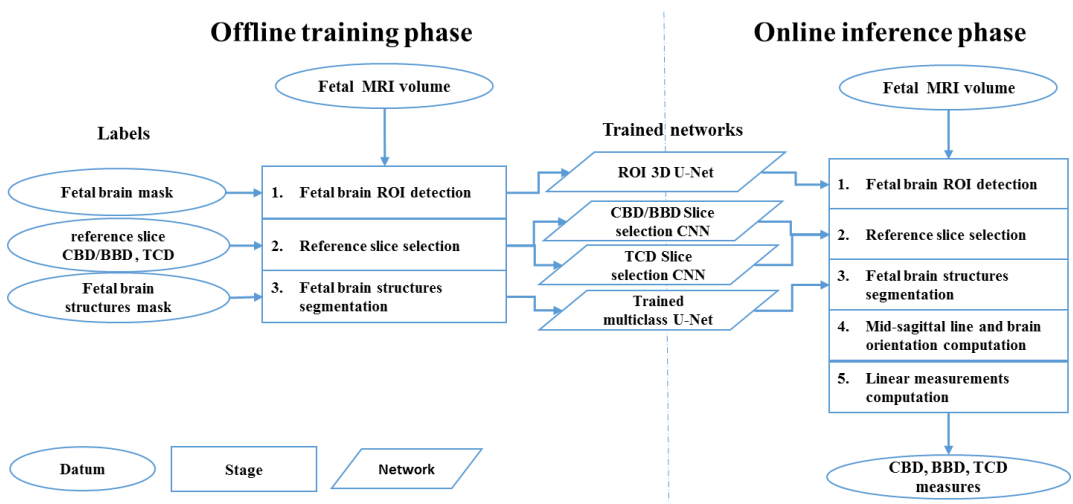

**Fig. 2**: Offline training (left) and online inference (right) phases. The offline training phase consists of the first three stages (rectangles). It inputs labeled data for each of the stages (ovals) and outputs four trained networks (parallelograms). The online inference phase uses these networks for classification followed by the last two stages.

### 3.1 Fetal brain ROI detection

The first stage computes the fetal brain ROI in the fetal MRI volume. The ROI is a 3D axis-aligned bounding box that contains the fetal brain. It is computed using a custom anisotropic 3D U-Net with a Dice loss function described in [8]. Briefly, the network inputs a ×4 downscaled version of the fetal MRI volume and outputs a coarse fetal brain voxel classification from which a tight bounding box is computed. This network achieves 100% ROI detection rate on a network trained with very few (~10) manually labeled volumes.

### 3.2 Reference slice selection

The next stage is the selection of the two reference slices on which the measurements are performed: one for the CBD/BBD, and another one for the TCD. The slices were selected with a CNN classifier trained for each type of reference slice using transfer learning in three steps: 1) fetal brain ROI pre-processing; 2) slice probability prediction of each slice in the fetal brain ROI, and; 3) reference slice selection based on the computed probabilities.

First, the fetal brain ROI, which is a grey-value matrix of size $h \times w \times n$ where $h$ and $w$ are the fetal brain bounding box height and width and $n$ is the number of slices, is resized to $s \times s \times n$ where $s = 1.5 \times \max(h, w)$ so that the slices are square. The volume is cropped by the resized ROI and resized to the size required by the ResNet50 network ($224 \times 224$). Note that only the ROI was changed, and not the volume itself, thus preserving its original aspect ratio and scale. The grey values are computed with bilinear interpolation and normalized to the [0,1] range. Next the probability of a slice to be the reference slice is computed with a modified version of the ResNet50 [24] CNN pre-trained on the ImageNet dataset. The two modifications are: 1) the last multi-class classification layer is replaced with a two-class softmax classification layer, and; 2) the same grey voxel values are input to each of the three RGB channels. Finally, the reference slice $k$ is selected as the slice with the highest probability.

The two networks are used for reference slice selection are trained and used for the inference in the same way. In the offline training phase, slice-wise image augmentations, e.g., random cropping, rotations, horizontal and vertical flip are applied on-the-fly to the fetal brain ROI slices in each epoch. To compensate for class imbalance, i.e. only one slice out of ~25 in each volume is a reference slice, a small subset $l < n - 1$ of non-reference slices are randomly selected as negative examples (in practice $l = 2$ yielded the best results). The network is trained with the Binary Cross Entropy loss function for 30 epochs. In the first 10 epochs, all layers are trained; in the next 20 epochs, only the last classification layer is trained.

### 3.3 Fetal brain structure segmentation

This stage performs multiclass semantic segmentation on all fetal brain MRI slices computed in the previous stage into four fetal brain components: cerebellum, left and right cerebrum hemispheres, and background. The segmentation is performed with 2D U-Net consisting of a Resnet34 encoder pre-trained on the ImageNet dataset.

In the offline training phase, slice-wise image augmentations, e.g., random horizontal and vertical flip, and brightness and contrast adjustment are applied on-the-fly to the fetal brain slices within ROI in each epoch. Brightness and contrast adjustments have been shown to improve unseen domain generalization in both MRI and in Ultrasound [25]. The brightness of pixel p is adjusted by an amount c with $brightness(p, c) = 1/\left(1 + e^{-(\log(p/1-p) + \log(c/1-c))}\right)$ in the range $[0.4, 0.8]$. The contrast of pixel $p$ is adjusted by an amount $c$ with $contrast(p, c) = 1/1 + e^{-(\log(p/1-p) \times c)}$ in the range $[0.4, 1.6]$. The network is trained with the Lovasz loss function [26] for 24 epochs. In the first 12 epochs, only the decoder layers are trained; in the next 12 epochs, both the encoder and decoder layers are trained. In the online inference phase, post-processing is applied to each slice output segmentation by nearest-neighbor interpolation followed by zero-padding (background class) to obtain the original slice size ($h \times w$).

### 3.4 Mid-sagittal line and brain orientation computation

The MSL and the brain orientation are computed from the ROI and the fetal brain structure segmentation.

The MSL is computed as the minimal margin line that separates the left and right fetal cerebral hemispheres with a Support Vector Machine (SVM) classifier with a linear kernel:

$$\left[\frac{1}{n}\sum_{i=1}^{n}\max(0, 1 - y_i \cdot (\mathbf{w}.x_i - b))\right] + \lambda\|\mathbf{w}\|^2$$

where $x_i$ is the pixel coordinates vector, $\mathbf{w} = (w_0, w_1)$ is the linear kernel weights vector, $y_i$ is the cerebral hemisphere index (–1 left, +1 right), $\lambda$ is a predefined regularization parameter, and $b$ is the bias. The SVM solution yields the values of $w$ and $b$ from which the MSL equation is computed: $y = -\frac{w_0}{w_1}x - \frac{b}{w_1}$. The SVM is executed for up to $10^8$ iterations with $\lambda = 10$. When the SVM does not converge, the MSL cannot be computed and a warning is issued.

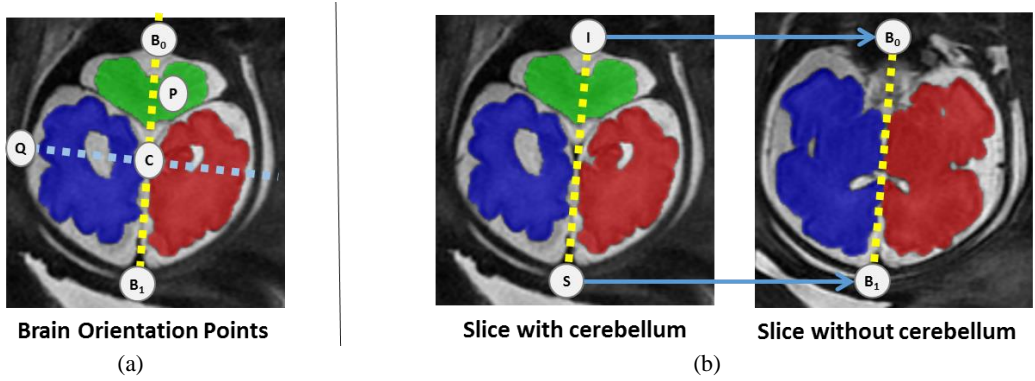

**Fig. 3**: Illustration of the mid-sagittal line (MSL) and brain orientation computation from the fetal brain structures: fetal cerebellum (green), left (blue) and right (red) fetal brain hemispheres: (a) the MSL (yellow) and its normal line (blue) passing through its middle point $C$ computed from points $B_0$ and $B_1$, the two intersection points of the MSL and the fetal brain ROI; $Q$ is the intersection between the MSL normal and the ROI boundary, $P$ is an arbitrary point inside the cerebellum; (b) two fetal MRI slices with and without the cerebellum; $B_0$ and $B_1$ are the closest inferior (I) and superior (S) points in the slice without the cerebellum.

The brain orientation, i.e. inferior/superior, is directly computed from the anatomical location of the cerebellum, which is inferior to the cerebral hemispheres (Fig. 3a). The MSL intersects the fetal brain ROI at points $B_0$ and $B_1$, from which the midpoint $C$ is computed. The line $QC$ is normal to the MSL that passes through C and intersects the fetal brain ROI at point $Q$. Next, an arbitrary point $P$ inside the cerebellum is sampled and classified with respect to the sign of the cross-product $QC \times P$. Since the cerebellum is inferior to the brain hemispheres, all points whose cross product sign are positive/negative are in the inferior/superior part of the brain. This computation is performed on all the slices that contain the cerebellum and then applied to the slices without the cerebellum by computing the Euclidean nearest neighbor distance in the slice plane (Fig. 3b)**.** This yields a mid-sagittal line for each slice in the fetal MRI volume.

### 3.5 Linear CBD, BBD and TCD measurements computation

The final stage computes the CBD, BBD, and TCD measurements with a geometric method akin to that used by expert radiologists.

The CBD measurement is computed in the reference slice from the MSL, the brain orientation and brain structures segmentation (Fig. 4a,b,c). First, the cerebrum width profile perpendicular to the MSL is computed from the cerebral brain segmentation boundary. Next, the Sylvian Fissure location is computed by finding the local minima of width profile that is the closest and superior to the brain mass center in the MSL. The CBD is the maximal width of the cerebral hemispheres superior to the Sylvian Fissure and perpendicular to the MSL.

The BBD measurement is computed by extending the CBD line to the skull contour on the same reference slice (Fig. 4c,d). First, the intensity derivative along the extended line is computed. Second, the local maxima of the derivatives is detected. Next, the inner skull contour pixels are identified by selecting the point with the maximum value from the two local extrema closest to the segmented cerebral brain boundary above a predefined threshold. The threshold value is used to filter out MR scanning imaging artifacts on the CSF, which appear as dark lines or spots, and therefore may cause noise when analyzing the intensity extrema.

The TCD measurement is defined as the maximal diameter of cerebellum contour convex hull of the fetal brain segmentation on the reference slice (Fig. 5).

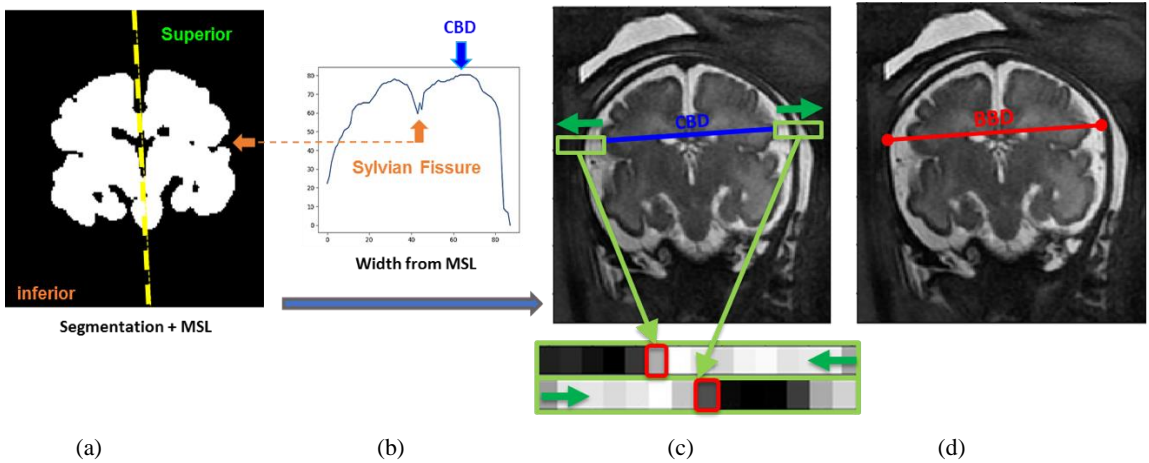

**Fig. 4**: Illustration of the CBD and BBD measurements computation: (a) fetal brain segmentation with the mid-sagittal line (yellow) superimposed; the orange arrows show the Sylvian Fissure; (b) the fetal brain width (vertical axis) as a function of the mid-sagittal line (horizontal axis); the blue arrow shows the local maxima, identifying the CBD; (c) the CBD (blue line) and its extension towards the fetal skull (green box) superimposed on the reference slice. The intensity profile along the CBD line (green box); red boundary pixels mark the locations of the inner fetal skull boundary; (d) BBD measurement (red line).

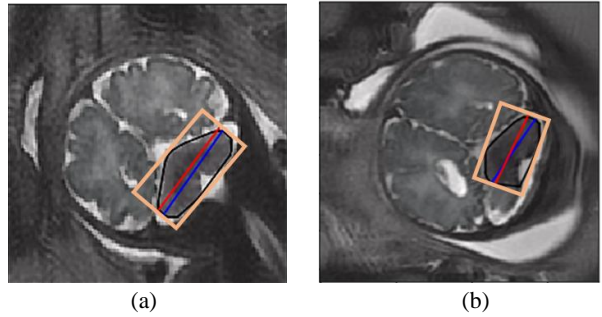

**Fig. 5**: Two examples of TCD measurements: reference slice, cerebellum convex hull (black contour), convex hull diameter (blue line), long axis bounding box (peach box) and its diameter (red line); (a) line angles agreement; (b) line angles disagreement.

### 3.6 Computation reliability estimation

Each step in the pipeline includes an automatic evaluation of its reliability. When no warnings are issued, the CBD, BBD and TCD measurement values are deemed accurate and trustworthy. When a warning is issued in one or more stages, the radiologist can inspect the result, make manual corrections as appropriate, or disregard the results. This reliability estimation may facilitate the use of the proposed method in a clinical environment.

Computation reliability warnings are issued for: 1) unreliable reference slice selection (stage 2), when the probability of the selected slice is below a predefined threshold (the preset value is 0.5, determined empirically). This heuristic is based on the observation that the slice selection network tends to be underconfident; 2) unreliable fetal brain structure segmentation (stage 3) and/or fetal brain orientation (stage 4), when brain orientations for five random points sampled on the cerebellum differ (our experimental results shows that five points are sufficient). This heuristic is based on the assumption that the cerebellum is inferior to the brain hemispheres, therefore an inconsistency in brain orientation may suggest that cerebellum segmentation is incorrect. Note that this heuristic is targeted to identify the cases when the fetal brain segmentation causes the failure the mid-sagittal line or fetal brain orientation, and is not designed to detect all possible segmentation errors; 3) unreliable mid-sagittal line (stage 4), when the mid-sagittal line angles of adjacent slices differ; 4) unreliable BBD measurement (stage 5) when the measurements on the original and CLAHE-enhanced [27] reference slices differ. The latter is determined by computing the measurement values on the reference slice with contrast limited adaptive histogram equalization (CLAHE) with tile size of 20×20 and clipping limit of 0.01. The rationale for this approach is that the fetal cerebrospinal fluid (CSF) might yield intensity inhomogeneity and imaging artifacts, so enhancing and equalizing the contrast may enhance the borders between CSF and brain parenchyma; 5) unreliable TCD measurement (stage 5), when the line angles between two methods of measurement differ more than 10º (pre-set value determined empirically): (a) the cerebellum convex hull diameter; and (b) the cerebellum bounding box long axis. Fig. 5b shows an example of unreliable TCD measurement detection. This heuristic is based on the characteristic butterfly-like shape of the fetal cerebellum in the coronal slice. When the cerebellum segmentation is incorrect, the symmetry may be affected, causing the line misalignment.

Note that the reliability of the computation of the fetal brain ROI (stage 1) is described in our previous paper [8] and has already been validated there. The fetal brain ROI computation produced correct results on all the datasets of our study.

## 4. Experimental results

To evaluate our method, we collected fetal brain MRI volumes, annotated them, and conducted six experimental studies.

### 4.1 Data

Anonymized fetal MRI studies were retrospectively obtained from the Tel Aviv Sourasky Medical Center, Tel Aviv, Israel, from a database of pregnant women referred for fetal MRI as part of routine clinical fetal assessment during 2012-2016. The full dataset consisted of 214 fetal brain MRI volumes (5,347 slices) of 154 singleton pregnancies (cases). Of these, 113 volumes (87 cases) were diagnosed as normal, and 101 volumes (67 cases) as abnormal with minor/moderate brain pathologies. Mean gestational age was 32 weeks (std=2.8, range 22-38, all weeks represented except 25). The MRI volumes of T2-weighted FRFSE sequence were acquired on a 1.5T General Electric Discovery MR450 scanner. Each volume consists of a mean 24 coronal slices (range 14-36) with mean in-plane resolution of 0.75×0.75 mm$^2$ (range 0.60–0.87×0.60–0.87 mm$^2$) and mean slice thickness of 4 mm (range 3–8). The *full dataset* was partitioned into the *training dataset* (164 volumes, 121 cases) with mean gestational age of 32 weeks (std=2.8, range 22-38). Of those, 68 cases were diagnosed as normal and 53 cases were diagnosed as abnormal. The mean gestational age of the *test dataset* (50 volumes, 33 cases) was 32 weeks (std=2.9, range 23-38). Of those, 17 cases were diagnosed as normal and 16 cases were diagnosed as abnormal. The training and test datasets are disjoint in terms of volumes and cases.

### 4.2 Dataset annotation

Manual reference slice selections and CBD, BBD and TCD measurements were obtained for all volumes in the full dataset volumes with ITK-SNAP [28] by the senior pediatric neuro-radiologist co-author (LBS). The mean time required for reference slice selection and manual measurement for all three measurements per volume was 110 secs (range 60-150 secs). Validated slice-based fetal brain structure segmentations were obtained on a subset of slices from the full dataset. The initial fetal brain segmentation was obtained with the method in [8] for 63 volumes (1,389 slices) from the training dataset. The resulting segmentations were post-processed by removing small connected components, selecting the 2-3 largest connected components (left and right fetal brain hemispheres and the cerebellum when present in the slice) and performing spectral clustering with discretization [29]. Of the resulting segmentations, 1,108 slices were reviewed and approved by a knowledgeable co-author (OBZ, a graduate student who has learned from the expert radiologist how to perform this task).

### 4.3 Evaluation metrics

We used the following metrics to quantify and compare the accuracy and variability of the annotations. Linear measurement differences were defined as $diff(l_1, l_2) = |l_1 - l_2|$ where $l_1, l_2$ are two linear measurement values. Slice selection differences were defined as $slice\_diff(s_1, s_2) = |s_1 - s_2|$ where $s_1, s_2$ are two selected slice indices. The slice selection accuracy was defined as $slice\_selection\_accuracy(s_1, s_2) = 1 - |s_1 - s_2|/N$ where N is the number of volume slices. The MSL angle difference was defined as $MSL\_diff\_Angle(l_1, l_2) = |angle(msl_1) - angle(msl_2)|$ where $msl_1, msl_2$ are two MSLs measured on the same volume. We used the Bland Altman method [30] to estimate the agreement between two sets of measurements. Agreement was defined by the 95% confidence interval, $CI_{95}$. For two sets, $CI_{95} = 1.96 * std_j(M_j^1 - M_j^2)$ where $M_j^1, M_j^2$, are the measurement of sets 1 and 2, each with $n$ measurements. The bias is $Bias = \frac{1}{n}\sum_{j=1}^{n}(M_j^1 - M_j^2)$.

### 4.4 Interobserver variability

The interobserver variability for the three manual linear measurements was established for subset from *the training set* (*n*=45) by computing the bias, difference and agreement metrics of the CBD, BBD and TCD measurements between two expert radiologists (co-authors EM and LBS). Table 1 shows the results.

| Measurements (in mm) | CBD | | | BBD | | | TCD | | |
|---|---|---|---|---|---|---|---|---|---|
| | Bias | $CI_{95}$ | Diff | Bias | $CI_{95}$ | Diff | Bias | $CI_{95}$ | Diff |
| Interobserver variability | 0.03 | 4.12 | 1.60 | -0.09 | 3.18 | 1.27 | 0.26 | 2.39 | 0.97 |

**Table 1:** Observer variability for the CBD, BBD and TCD measurements between two expert radiologists: Bias, 95% confidence interval, and difference.

### 4.5 Studies and results

The first study evaluated the overall performance of the method and of the self-assessment of reliability. The next four studies evaluated the performance of the various steps of the method and justified the algorithmic choices. The last study evaluated the performance of the method with gestational age and fetal brain abnormalities. The entire pipeline was tested on the *test dataset*, which is disjoint from the *training set*. The results for the entire pipeline for the *full dataset* are provided because our method includes both model-based and machine-learning methods.

The *training dataset* was divided into various subsets that were used for both training and validation of the five stages of the fetal biometric measurements pipeline. The size of the training and validation datasets for each stage was determined according to the specific characteristics of the step. For the fetal brain ROI detection, the pretrained network in [8] was trained with 6 volumes and validated with 29 volumes. For the reference slice selection, 49 volumes were used for training and 43 for validation for both the CBD/BBD and the TCD slice selection networks. For the fetal brain structure segmentation, 53 volumes were used for training and 10 for validation. For the linear measurements computation, which includes the fetal brain orientation computation, and the computation reliability estimation, 45 volumes were used for the parameters value selection. The training, validation, and test datasets are disjoint for all experiments.

**1. Biometric measurement evaluation.** This study evaluated the accuracy of the computed CBD, BBD and TCD measurements and of the self-assessment of reliability. Table 2 summarizes study results.

| **Measurements (in mm)** | **CBD** | | | **BBD** | | | **TCD** | | |
|---|---|---|---|---|---|---|---|---|---|
| | **Bias** | $CI_{95}$ | **Diff** | **Bias** | $CI_{95}$ | **Diff** | **Bias** | $CI_{95}$ | **Diff** |
| **Test dataset (*N*=50)** | -0.08 | 3.94 | 1.48 | -0.43 | 3.26 | 1.21 | -0.82 | 3.27 | 1.26 |
| **Test dataset, reliable cases CBD (*N*=47), BBD (*N*=46), TCD (*N*=40)** | 0.70 | 3.09 | 1.44 | -0.03 | 2.84 | 1.10 | -0.26 | 2.17 | 0.83 |
| **Full dataset (*N*=214)** | -0.32 | 4.32 | 1.55 | -0.47 | 3.97 | 1.47 | -1.14 | 3.00 | 1.40 |
| **Full dataset, reliable cases CBD (*N*=206), BBD (*N*=203), TCD (*N*=183)** | 0.69 | 3.92 | 1.52 | 0.19 | 3.98 | 1.47 | -0.63 | 2.25 | 1.06 |

**Table 2**: Accuracy of computed CBD, BBD and TCD measurements with and without excluding the volumes identified as unreliable by the reliability self-assessment (reliable cases) on the test and full datasets. N indicates the number of volumes for each measurement.

All results are below the inter-observer variability range. The reliability self-assessment on the full dataset identified uncertainty in 28 volumes out of 214 (13%) for the TCD measurement, 8 volumes out of 214 (3%) for BBD/CBD slice selection, 3 volumes out of 214 (1.5%) for TCD slice selection, 3 volumes out of 214 (1.5%) for the BBD measurement, and no unreliable volumes for the brain orientation and the mid-sagittal line angle computation. The reliability self-assessment method also improved the variability measures on all datasets. For the *test dataset*, the $CI_{95}$ decreased from 3.94mm to 3.09mm (27%), from 3.26mm to 2.84mm (15%) and from 3.27mm to 2.17mm (50%) for the CBD, BBD and the TCD measurements respectively.

**2. Reference slice selection: network selection.** This study compared the accuracy of five different networks that were used for reference slice selection: ResNet18, ResNet34, ResNet50, DenseNet121[31] and VGG16 [32] pre-trained on ImageNet. For each, we trained the network with its recommended hyperparameters on 1,130 slices from 49 volumes and validated them with different training and validation subsets of the *training dataset* on the TCD reference slice selection (Table 3). ResNet50 has the highest accuracy on the validation split with no significant difference in accuracy between all ResNets and DenseNet.

| **TCD reference slice selection accuracy** | **Network** | | | | |
|---|---|---|---|---|---|
| | **ResNet18** | **ResNet34** | **ResNet50** | **DenseNet121** | **VGG16** |
| **Training (*N*=49)** | 0.961 | 0.965 | 0.968 | 0.969 | 0.960 |
| **Validation (*N*=17)** | 0.978 | 0.977 | **0.978** | 0.977 | 0.963 |

**Table 3**: Accuracy of five reference slice selection networks for TCD reference slice selection task for the training and validation splits. The selected network results are shown in **bold**.

We then trained the ResNet50 network for CBD/BBD and TCD reference slice selection on different training and validation subsets of the *training dataset*. We evaluated the reference slice selection accuracy with respect to the manual reference slices selected on the *test dataset* (Table 4). All networks achieved a mean difference of < 1 slice.

| Reference slice selection | Split | Reference slice difference # | | Reference slice selection accuracy |
|---|---|---|---|---|
| | | Mean | Maximum | Mean |
| BBD/CBD | Training (*N*=49) | 0.454 | 2 | 0.979 |
| | Validation (*N*=43) | 0.411 | 3 | 0.981 |
| | Test (*N*=50) | 0.795 | 3 | 0.968 |
| TCD | Training (*N*=49) | 0.477 | 3 | 0.981 |
| | Validation (*N*=43) | 0.457 | 1 | 0.977 |
| | Test (*N*=50) | 0.604 | 2 | 0.975 |

**Table 4**: Reference slice selection mean and maximum # of slices and mean accuracy on the BBD/CBD and TCD reference slice for ResNet50 for the training, validation, and test datasets.

**3. Fetal brain structure segmentation accuracy.** This study evaluated the accuracy of the fetal brain structure segmentation on the *full dataset*. We used splits on the validated dataset of 1,008 (train) and 100 (test) slices each. The mean Dice coefficient on the training and test splits was 0.942 and 0.944 respectively. The Dice coefficient on these splits was 0.942 and 0.945 for the right hemisphere, 0.940 and 0.946 for the left hemisphere, and 0.849 and 0.851 for the cerebellum. This indicates that the segmentations are accurate and reliable.

**4. Mid-sagittal line angle accuracy.** The MSL angle accuracy of the computed and the manual measurement on the *full dataset*, $CI_{95}$ is 4.86°; the mean difference is 1.93°.

**5. Computation reliability estimation validation.** This study evaluated the reliability estimation mechanism on the *test dataset*. An experienced radiologist visually inspected all the 50 volumes of the *test dataset* and graded each of the three measurements, CBD, BBD and TCD, as valid or unacceptable. The selection results were then compared to the computation reliability warnings that were automatically issued by the method.

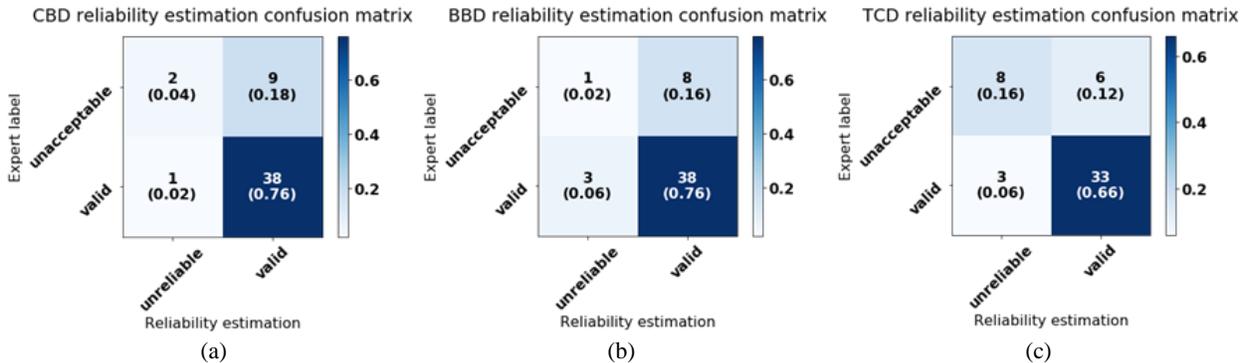

(a)      (b)      (c)

**Fig. 6:** Confusion matrices of the reliability estimation computation: manual expert radiologist evaluation (vertical axis) vs. automatic reliability estimation: (a) CBD; (b) BBD; (c) TCD measurement.

Fig. 6 shows the confusion matrices for each measurement. The accuracy of the reliability estimation computation is 80%, 78% and 82% for CBD, BBD, and TCD measurements, respectively. Note that the expert accepted 78%, 82% and 74% of the CBD, BBD and TCD measurements, respectively. For the cases where the expert did not accept and the method did accept, the three measurement variability was 2.95mm, 1.65mm and 1.92mm, respectively, which is lower than the inter-observer variability. This means that the computed measurement is clinically acceptable. Upon inspection, the main reason for which the method did not rule out these cases is the measurement angle, which was not covered by our heuristics.

6. **Performance with gestational age and abnormalities.** This study examines the impact of the fetus gestational age (GA) and fetal brain abnormalities on the entire pipeline on the *full dataset*, after elimination by the reliability estimation.

First, we compute the mean deviation ratio from the absolute value and its standard deviation for each measure for each GA. Fig. 7 shows the results. The mean error as a function of GA is consistently the same across all measures, except for the lower GA in the BBD measurements -- week 24, with a 4% deviation vs. 2% for all other GAs -- and TCD measurements -- week 22, with a 6% deviation vs. 3% for all other GAs. However, this discrepancy may be caused by the fact that we have a single sample for this GA (week 22).

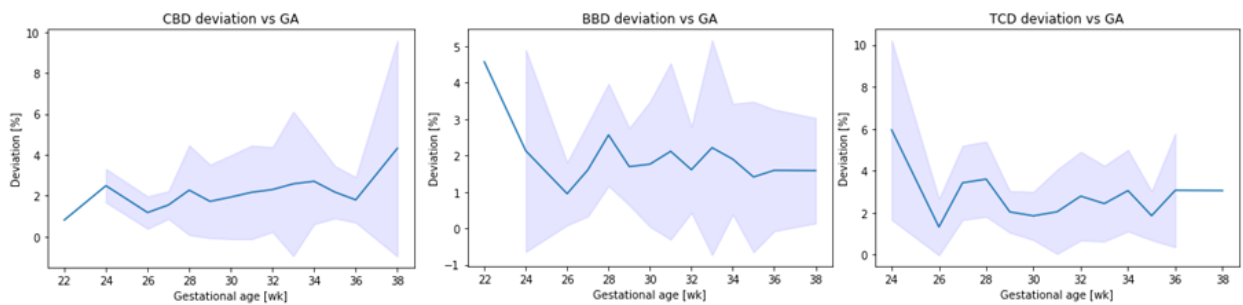

**Fig. 7:** Three graphs showing the CBD, BBD, TCD measurements mean deviation (%, vertical axis) as a function of the Gestational Age (weeks, horizontal axis) for the *test dataset*.

Second, we analyze the impact of fetal abnormalities by computing the joint distribution of the error and fetal GA vs. case diagnosis (normal vs. abnormal). Fig. 8 shows the results. Note that the marginal distributions of errors for normal and abnormal cases over all measurements are similar. This means that the subject condition does not affect the system performance. Further investigation shows that the specific anomalous cases that have large deviation in CBD/BBD measurements are from the Malformation of Cortical Development (MCD) diagnosis. The MCD group shows high variability of brain disorders which can affect the measurements. However, cases with Dolicocephaly and extra axial fluid diagnoses, which are the ones that should be detected with CBD/BBD measurements, performed well with a measurement deviation of < 2mm even though there is no representation for these groups in the training dataset.

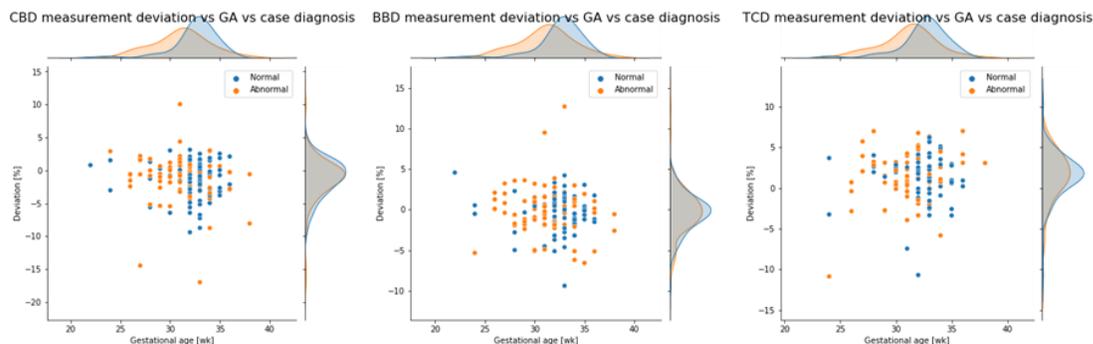

**Fig. 8:** Three scatter plots showing the normal (blue dot) and abnormal (orange dot) cases of the CBD, BBD, TCD measurements mean deviation (%, vertical axis) as a function of Gestational Age (weeks, horizontal axis) for the *full dataset*. The two graphs on top and on the right shows the Gestational Age and measurements mean deviation distribution of the normal (blue) and abnormal (orange), respectively.

5. **Conclusions**

To the best of our knowledge, this is the first fully automatic method that measures biometric linear measurements of the fetal brain from MRI according to accepted clinical guidelines. It is based on a hybrid approach combining deep learning methods and geometrical algorithms for fetal brain ROI detection, fetal brain component segmentation, MSL and brain orientation estimation and three main biometric linear measurements. Two unique features of our algorithm are a new reference slice selection method using CNNs, and a new method for self-assessment of reliability to alert clinicians when the computed measurements may be unreliable.

We believe that the methodology and experimental results presented here can be useful for developing methods for automatically computing biometric linear measurements in volumetric scans. The pipeline is generic in its first three stages and in its approach to reliability self-evaluation. The deep learning methods used rely on a few dozen annotated datasets, which makes it practical.

The proposed method achieves human-level performance, while handling high input variability represented in clinical use: a variety of gestational ages, pathological fetal brain conditions, and diverse MRI scanning parameters. It therefore may be useful in the assessment of fetal brain biometry, and improving routine clinical practice.

**Declarations**

**Funding:** This research was supported by Kamin grant 63418 from the Israel Innovation Authority; Joseph Bar Nathan Trustee of Mychor Trust Fund.

**Human rights statement:** The included human study has been approved by the Tel Aviv Sourasky Medical Center institutional review board # 02-001 and performed in accordance with ethical standards.

**Conflict of interest:** None of the authors has any conflict of interest.

**Informed consent:** Informed consent was applied.

**Presentation at conferences:** an earlier, partial version of this research was presented as an abstract at CARS 2020.